\documentclass[fleqn,12pt]{article}
\usepackage{graphicx}

\begin{document}

\title{Coupled collective motion in nuclear reactions}
\author{Ph. Chomaz and C. Simenel \\
GANIL, BP 55027, 14076\ CAEN\ cedex 5 France}
\maketitle

\begin{abstract}
In this paper, we review the roles of collective modes in nuclear reactions.
We emphaize the strong couplings of various collective states with the
monopole and quadrupole motions. In inelastic excitation, these couplings
can be seen as an important source of anharmonicity in the multiphonon
spectrum. In fusion reaction, the breathing and quadrupole motions strongly
affects the oscillation of protons against neutrons. Finally, the
modification of the collective properties induced by a large amplitude
dilution might be the origin of the nuclear multifragmentation, directly
related to the nuclear liquid-gas phase transition. In the three cases we
derive a coupling matrix element which appears to be in good agreement.
\end{abstract}

\section{INTRODUCTION}

Strongly interacting systems with many degrees of freedom are the prototypes
of complex systems. As a consequence of this complexity, their dynamics is
expected to present disorder or chaos. However, in reaction to an external
stress, such systems appear to often self-organize in simple collective
motions. Of particular importance is the occurrence of collective vibrations
which, in general, are surprisingly harmonic despite their chaotic
environment.

This paradox is well illustrated by the atomic nucleus. Indeed, on the one
hand, following the Bohr ideas, the compound nucleus resonances sign the
occurrence of quantum chaos already just above the neutron threshold \cite
{BM2}. On the other hand, in the same excitation energy domain, the nucleus
is known to exhibit a large variety of collective vibrations (called
phonons) \cite{Har01}. The first quantum of oscillation is associated with
the giant resonances, anomalously large cross sections observed in some
nuclear reactions. The first one to be discovered was the giant dipole
resonance (GDR) \cite{Ba47} interpreted as a collective oscillation of the
neutrons against the protons \cite{GT48}. Then came the giant quadrupole
resonance (GQR), an oscillation of the nucleus shape between prolate to
oblate deformations \cite{fuk72}, and the giant monopole resonance (GMR) or
breathing mode, an alternation of compressions and decompressions of the
whole nucleus \cite{Ma76a}. Since then many other resonances have been
uncovered \cite{Har01,Wo87}. Twenty years ago, giant resonances have also
been observed in hot nuclei up to several-MeV temperature \cite{Ne83}. This
demonstrates the survival of ordered vibrations in very excited system which
are known to be chaotic.

The study of this amazing self-organization of the nucleus in collective
vibrations and its transition from order to chaos is one of the important
subjects in modern nuclear physics.

In this article, we will focus on the onset of disorder through the coupling
between various modes. First we will show that collective vibrations induce
monopole (GMR) and quadrupole (GQR) oscillations.\ This means that the
coupling matrix elements between a quantum of vibration and a state with a
GMR or a GQR built on top of it are large. We will present results from two
''orthogonal'' approaches, i.e. boson mapping (BM) and time-dependent
mean-field (TDMF), showing the same effect. Then we will move to larger
amplitude motion reached in reactions. First we will discuss the effect of
the large amplitude monopole and quadrupole oscillations induced by fusion
reaction on the GDR. Then, we will move to even more violent reactions for
which a rapid expansion of the produced nuclear system have been observed
and interpreted as the result of fast decompression of the matter. We will
show how this large-amplitude breathing mode affects all the other
collective states, which may even become unstable. We will also make the
bridge between this coupling of collective motions and the liquid-gas phase
transition.

Finally we will connect the three studied phenomena with a non linear
coupling between a vibration and a GMR or GQR built on top of it.

\section{Theoretical framework \protect\cite{RS81}}

In quantum mechanics, harmonic oscillations are associated with boson
degrees of freedom. From the microscopic point of view, these bosons can be
understood as being built from fermion pairs, which carry boson quantum
numbers. However, the number of possible pairs must be large enough to
insure that the effects of the fermion antisymmetrization do not introduce
significant deviations from a boson behavior. In particular, the excitations
of small fermionic systems are not expected to be well described by a boson
picture, because the Pauli exclusion principle imposes constraints that
cannot be easily accounted for in a boson representation.  From a formal
point of view the relation between fermion pairs and bosons can be
explicitly worked out using boson mapping techniques \cite{RS81}. We will
use one of these methods in the first study we are presenting.

Fermionic approaches can also be followed. For example, giant resonances are
often described using time dependent mean field approaches like the Time
Dependent Hartree-Fock approximation (TDHF). Indeed, they correspond to the
response of the system to an external (collective) one-body field and
mean-field approaches are tailored to take care of such excitations.
Moreover, giant resonances directly affect the time evolution of one-body
(collective) observables which are well predicted by mean-field approaches.
The small amplitude reduction of these approaches is equivalent to the
random phase approximation (RPA) \cite{RS81}. Being a linearization of the
equation of motion, it corresponds to a harmonic picture. However, since the
mean-field depends upon the actual excitation, TDHF is a non linear theory
and hence contains couplings between collective modes. For example the
quadratic response takes into account the couplings between one and two
phonon states coming from the 3-particle 1-hole and 1-particle 3-hole
residual interaction. In fact TDHF is optimized for the prediction of the
average value of one body observables. Through non-linearities and time
dependence, it takes into account the effects of the residual interaction as
soon as the considered phenomenon can be observed in the time evolution of a
one body observable. This was already the case for the RPA, which through
the time dependence takes into account the particle-hole residual
interaction and goes beyond the static mean field which is limited to the
hole-hole terms. In this article, we will go beyond the RPA treatment either
working out the quadratic response to TDHF or directly performing full TDHF
calculations \cite{kim}.

\section{\protect\smallskip Multiphonon anharmonicities due to the coupling
with GMR and GQR}

\smallskip Let us start with a direct manifestation of the coupling between
collective states: the anharmonicity of multiphonon spectra.

\subsection{\protect\smallskip TDHF picture}

\begin{figure}[tbh]
\begin{center}
\includegraphics[width=8cm]{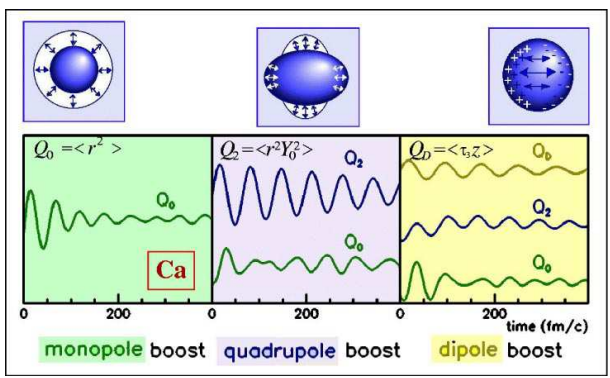}
\end{center}
\caption{Evolutions of the monopole, quadrupole and dipole moments (solid
lines) as a function of time for monopole (a), quadrupole (b) and dipole (c)
excitations in $^{40}$Ca. }
\label{fig:1}
\end{figure}

Time dependent approaches provide an intuitive understanding of collective
motions \cite{cho-Landau}. Let us for example look at the TDHF dynamics for
the $^{40}$Ca nucleus which has been initially perturbed by a collective
boost. For this simulation we have used the TDHF\ 3D code developed by P.
Bonche and coworkers \cite{kim} with the SGII Skyrme force \cite{sg}.

In Fig. \ref{fig:1}, we followed the monopole, quadrupole and dipole
response for three initial conditions:

\begin{itemize}
\item  A monopole boost using ${Q}_{0}=\frac{1}{\sqrt{4\pi }}\sum_{i}({r}%
_{i}^{2}-\langle {r}_{i}^{2}\rangle (t=0))$ as a boost generator. Because of
the spherical symmetry, a monopole boost can only trigger monopole modes.
Therefore, we only observe $\langle {Q}_{0}\rangle (t).$

\item  A quadrupole boost generated by ${Q}_{2}=\sum_{i}{r}%
_{i}^{2}Y_{0}^{2}\left( {\theta }_{i},{\varphi }_{i}\right) .$ The parity
conservation forbids any dipole excitation when a quadrupole velocity field
is applied to a spherical nucleus. Conversely, breathing modes can be
triggered by the quadrupole oscillation so that we do follow both the
quadrupole $\langle {Q}_{2}\rangle (t)$ and the monopole $\langle {Q}%
_{0}\rangle (t)$ responses.

\item  An isovector dipole boost induced by ${Q}_{D}=Z/A\sum_{n}z_{n}-N/A%
\sum_{p}z_{p}.$ This excitation can be both coupled to the quadrupole and
monopole oscillations so that we monitor the three moments, $\langle {Q}%
_{0}\rangle (t)$, $\langle {Q}_{2}\rangle (t)$ and $\langle {Q}_{D}\rangle
(t).$
\end{itemize}

\subsubsection{\protect\smallskip Linear response and collective states}

In Fig. \ref{fig:1}, we observe that the collective boosts induce
oscillations of the associated moments as expected from the RPA picture.
They are only slightly damped in the GQR and GDR cases (fig. 1-b and 1-c
respectively) while in the GMR case (fig. 1-a) beatings, characteristic of a
Landau damping \cite{cho-Landau}, are observed. This means that the dipole
and quadrupole strengths are mostly concentrated in a single resonance while
the monopole one is fragmented.

\begin{figure}[tbh]
\begin{center}
\includegraphics[width=8cm]{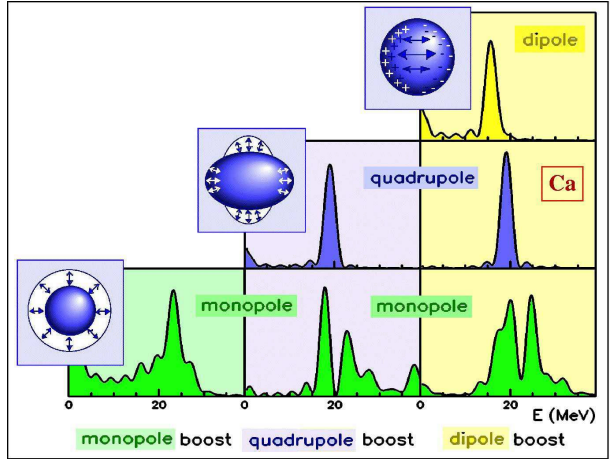}
\end{center}
\caption{Monopole, quadrupole and dipole spectra obtained through the
Fourier transform of the time dependent response for a monopole, quadrupole
and dipole excitation in the $^{40}$Ca.}
\label{fig:2}
\end{figure}

Looking at the amplitude of the first oscillation $\langle{Q}_{\nu
}\rangle_{max}$ as a function of boost strength ($k_{\nu }$) confirms the
linearity of this response \cite{cedric-phonon}. To get a deeper insight
into the response, we study the Fourier transform $F_\nu (\omega )$ of $%
\langle{Q}_{\nu }\rangle(t)/k_{\nu }$ which is nothing but the RPA response
when the velocity field $k_{\nu }$ is small enough to be in the linear
regime. We see in Fig. \ref{fig:2} that the dipole and quadrupole modes are
concentrated in a unique peak. The monopole is fragmented but the various
peaks are in the same energy region so that they can be approximated by a
single mode with a large Landau width.

\subsubsection{Quadratic response}

In Fig. \ref{fig:1}, one can see that large amplitude dipole (fig. 1-c) and
quadrupole (fig. 1-b) motion induces variations of the nucleus radius $%
\langle{Q} _{0}\rangle.$ Looking at the matter distribution one can see that
the central density $\rho _{0}$ is affected by the dipole or quadrupole
motion. Since the central density can be modified only by monopole states
this imposes that the large amplitude motion gets coupled with such
breathing modes. In the same way a large amplitude dipole oscillation
induces a quadrupole deformation of the nuclear potential and so gets
coupled with the GQR. These observations lead to the conclusion that we are
in the presence of a non-linear excitation of a giant resonance (GMR or GQR,
generically called $\mu $ in the following) on top of the collective motion
(GQR or GDR, generically called $\nu $ in the following) initially excited
through the collective boost ${Q}_{\nu }$. As expected from the quadratic
response theory \cite{cedric-phonon}, the amplitude of the first oscillation
of $\langle{Q}_{\mu }\rangle(t)$ is quadratic in the excitation velocity $
k_{\nu }$ and its time dependence corresponds to $\left( \cos (\omega _{\mu
}t)-1\right) $, i.e. it is initially in phase quadrature with $\langle{Q}%
_{\nu }\rangle(t)$ but it is oscillating with the frequency of the mode $\mu 
$ and not the one of the initially excited collective state $\nu $.

This interpretation is confirmed by the Fourier transform of $\langle{Q}%
_{\mu }\rangle(t) $ associated with the excitation of ${Q}_{\nu }$ (Fig. \ref
{fig:2}). Let us first start with the quadrupole strength non-linearly
excited by a dipole boost. The observed peak is identical to the
GQR-response. This is a clear indication that the observed state is indeed a
GQR built on top of the GDR. It should be noticed that this frequency is
different from the one of the underlying dipole motion. The monopole case is
more complex because of the presence of a strong Landau spreading and it
seems that the strengths of the various monopole states depend upon the
considered boost. This indicates that the coupling leading to the excitation
of an additional monopole state depends upon the collective mode initially
excited.

\subsubsection{Quadratic response and Couplings between states}

\begin{table}[tbp] \centering%
\begin{tabular}{lcccc}
\hline
{\ $|\nu \rangle$} & {\ $\omega _{\nu }$} & {\ $q_{\nu }$} & {\ $\langle\nu |%
{V}|\nu 0\rangle$} & {\ $\langle\nu |{V}|\nu 2\rangle$} \\ 
{\ } & {\ (MeV)} &  & {\ (MeV)} & {\ (MeV)} \\ \hline
{\ $|0\rangle_{^{40}Ca}$} & $22.9$ & $11.6$ & - & $0$ \\ 
{\ $|2\rangle_{^{40}Ca}$} & $18.6$ & $21.4$ & $-4.28$ & - \\ 
{\ $|D\rangle_{^{40}Ca}$} & $17.2$ & $3.47$ & $-4.58$ & $-3.92$ \\ \hline
{\ $|0\rangle_{^{208}Pb}$} & $15.7$ & $57.1$ & - & $0$ \\ 
{\ $|2\rangle_{^{208}Pb}$} & $11.1$ & $99.0$ & $-2.17$ & - \\ 
{\ $|D\rangle_{^{208}Pb}$} & $13.0$ & $8.94$ & $-2.40$ & $-0.70$ \\ \hline
\end{tabular}
\caption{Energies, transition probabilities $q_\nu$ and coupling
coefficients of the GMR, GDR and GQR in the $^{40}$Ca and $^{208}$Pb. $q_\nu$ is expressed in fm$^2$
 for the GMR and GQR
 and in fm for the GDR.\label{tdhf}} 
\end{table}%

Assuming for each multipolarity a unique state $|\mu \rangle$ non linearly
excited one can use the quadratic response theory \cite{cedric-phonon} to
extract the residual interaction matrix element $v_{\mu }$ between $|\nu
\rangle$ and $ |\nu \mu \rangle$ from the amplitudes of the induced
oscillations $\langle{Q}_{\mu }\rangle_{max}$ using 
\begin{equation}
v_{\mu }=\frac{\langle{Q}_{\mu }\rangle_{max}\omega _{\mu }}{2k_{\nu
}^{2}q_{\nu }^{2}q_{\mu }}  \label{coupl}
\end{equation}
where $q_{\eta }$ is the transition matrix element between the ground state $
|-\rangle$ and the collective state $|\eta \rangle,$ $q_{\eta
}=\langle-|Q_{\eta }|\eta \rangle,$ which can be derived from the linear
response with $\eta=\nu$ since 
\begin{equation}
q_{\nu }^{2}=\frac{\langle{Q}_{\nu }\rangle_{max}}{2k_{\nu }}.
\end{equation}
The results for the $^{40}$Ca and $^{208}$Pb are presented in table 1. The $
\omega _{\nu }$ are computed from the time to reach the first maximum of $%
\langle{Q }_{\nu }\rangle(t)$. The relative sign of $v_{\mu }$ and $q_{\mu }$
is given by the early evolution of the moments ${\langle{Q}_{\mu }\rangle}$
(see Eq. \ref{coupl}). They appear to be all negative. The couplings $v_{\mu
}$ are large of the order of few MeV. As we will see those findings are in
qualitative agreement with the results of ref. \cite{Mumu} which is using a
completely different approach.


\smallskip

\subsection{Comparison with boson mapping calculations}

\smallskip

\begin{table}[tbp] \centering%
\begin{tabular}{lcccrcc}
\hline
{$|\nu \rangle$} & $J^{\pi }~T$ & $E_{harm}$ & $EWSR$ & {$\langle\nu |{V}%
|\nu 0_{1}\rangle$} & {$\langle\nu |{V}|\nu 0_{2}\rangle$} & {$\langle\nu |{V%
}|\nu 2\rangle$} \\ 
&  & (MeV) & $(\%)$ & (MeV) & (MeV) & (MeV) \\ \hline
{$|0_{1}\rangle_{^{40}Ca}$} & $0^{+}~0$ & $18.25$ & $30$ & $-2.13$ & $-2.36$
& $-$ \\ 
{$|0_{2}\rangle_{^{40}Ca}$} & $0^{+}~0$ & $22.47$ & $54$ & $-2.03$ & $-3.96$
& $-$ \\ \hline
{$|1_{1}\rangle_{^{40}Ca}$} & $1^{-}~1$ & $17.78$ & $56$ & $-1.38$ & $-2.12$
& $-1.25$ \\ 
{$|1_{1}\rangle_{^{40}Ca}$} & $1^{-}~1$ & $22.03$ & $10$ & $-1.48$ & $-2.16$
& $+0.73$ \\ \hline
{$|2\rangle_{^{40}Ca}$} & $2^{+}~0$ & $16.91$ & $85$ & $-1.36$ & $-2.49$ & $%
-0.36$ \\ \hline
\end{tabular}
\caption{ RPA one-phonon basis for the nucleus $^{40}$Ca. 
For each state, spin, parity, isospin, energy and percentage 
of the EWSR are reported in addition to the coupling coefficients. 
\label{Ca40.rpa}} 
\end{table}%

\smallskip

\begin{table}[tbp] \centering%
\begin{tabular}{lccrccc}
\hline
{$|\nu \rangle$} & $J^{\pi }~T$ & $E$ & $EWSR$ & {$\langle\nu |{V}|\nu
0_{1}\rangle$} & {$\langle\nu |{V}|\nu 0_{2}\rangle$} & {$\langle\nu |{V}%
|\nu 2\rangle$} \\ 
&  & (MeV) & $(\%)$ & (MeV) & (MeV) & (MeV) \\ \hline
{$|0_{1}\rangle_{^{208}Pb}$} & $0^{+}~0$ & $13.61$ & $61$ & $-1.87$ & $-0.92$
& $-$ \\ 
{$|0_{2}\rangle_{^{208}Pb}$} & $0^{+}~0$ & $15.02$ & $28$ & $-1.32$ & $-1.16$
& $-$ \\ \hline
{$|1_{1}\rangle_{^{208}Pb}$} & $1^{-}~1$ & $12.43$ & $63$ & $-0.79$ & $-0.59$
& $-0.68$ \\ 
{$|1_{2}\rangle_{^{208}Pb}$} & $1^{-}~1$ & $16.66$ & $17$ & $~0.00$ & $~0.00$
& $-0.64$ \\ 
{$|2\rangle_{^{208}Pb}$} & $2^{+}~0$ & $11.60$ & $76$ & $-0.64$ & $-0.48$ & $%
-0.74$ \\ \hline
\end{tabular}
\caption{ Same as table for the nucleus $^{208}$Pb. \label{Pb208.rpa}} 
\end{table}%

\smallskip

\smallskip

\smallskip In ref. \cite{Mumu} a completely different approach is used to
infer the same matrix elements: the fermionic Hamiltonian is first mapped
into a bosonic one making a connection between any particle-hole excitation
and a boson. Because of the fermionic anticommutation relations (Pauli
principle) a particle-hole excitation operator is mapped into an infinite
series of boson operators. As a consequence, even if the fermionic
Hamiltonian is containing only two body interaction, the boson Hamiltonian
is a infinite series with many boson interactions not conserving the boson
number. To be manageable this series has to be truncated and, in the
application presented in \cite{Mumu}, only the terms containing up to four
boson creation or anihilation operators have been conserved. Then a RPA
transformation is applied, introducing collective phonons which optimizes a
harmonic picture to the system properties \cite{Beaumel}. Using those
collective degrees of freedom, the Hamiltonian then contains an harmonic
part plus various interactions. An important one is the coupling between
one- and two phonon states.

For practical reasons only the most collective phonons have been selected in
ref. \cite{Mumu} some of them are presented in table 2 and 3 for $^{40}$Ca
and $^{208}$Pb. To illustrate their degree of collectivity the part of the
energy weighted sum rule (EWSR) they are exhausting is given together with
their energy. 
Table 2 and 3 also give the coupling matrix elements between different
collective states and one of the two GMRs or the GQR built on top of them.

\smallskip From the quantitative point of view, the non-linear coupling
extracted from TDHF appears to be $50\%$ larger than the one reported in
Table 2 and 3 \cite{Mumu}. This is a reasonable agreement since TDHF re-sums
all the individual couplings as shown in \cite{cedric-phonon}. Summing the
contributions of the different collective states considered in ref. \cite
{Mumu}  reduces the difference between the reported values. However, the
phonon basis studied in ref. \cite{Mumu} being incomplete , it is expected
that the TDHF results remains higher. It should be also noticed that some
differences can remain due to the approximations involved in the different
approaches.

\subsection{\protect\smallskip Consequences on the multiphonon spectrum}

The couplings between phonon states can be used to derive the energies of
the various phonon states. It appears that the very large matrix elements
exciting a GMR or a GQR on top of any state are the main source of
anharmonicites. As far as the two-phonon states are concerned the coupling
with the three-phonon states coming from such a non-linear coupling induces
an energy shift of about $2$ MeV for Ca and $1/2$ MeV for Pb. This might be
the explanation of the phonon anharmonicity puzzle discussed by many authors 
\cite{cho1,eml,aum} and summarized in \cite{Mumu}.

\section{\protect\smallskip Pre-equilibrium GDR in fusion reactions}

\smallskip

It has been recently proposed that GR can play an essential role in fusion
reactions. In particular, fusion reactions with N/Z asymmetric nuclei may
lead to the excitation of a GDR because of the presence of a net dipole
moment in the entrance channel \cite{cho2}. Experimental indications of the
possible existence of such new phenomenon have been reported in refs. \cite
{fli,amo98,pie01} for fusion reactions and \cite{san,pap99} for deep
inelastic collisions.

\subsection{\protect\smallskip Collective dynamics in fusion.}

As an example \cite{PRL-cedric}, we have computed the central collision of $
^{20}O+^{20}Mg$ at 1 MeV per nucleon with the TDHF\ approach. The system
rapidly fuses producing an excited $^{40}$Ca nucleus. It presents a strong
quadrupole oscillation around a slowly damped deformation. Since $^{20}$O
has a N over Z ratio different from this of $^{20}$Mg (respectively 1.5 and
0.67), the center of mass of the protons is initially different from the
neutrons one. As the time goes on this dipole moment $Q_{d}$ (i.e. the
distance between the neutron and proton center of mass) oscillates (see Fig. 
$ 3$). To study the induced motion one can plot the dipole moment $Q_{_{d}}\ 
$ as a function of the velocity of protons against neutrons which can be
considered as its conjugated moment, $P_{_{d}}$. We observe a spiral in this
collective phase space $(Q_{_{d}},P_{_{d}})$ , i.e. oscillations in phase
quadrature of the conjugated dipole variables. This is a clear signal of the
presence of a damped collective vibration.

\begin{figure}[tbp]
\begin{center}
\includegraphics[width=8cm]{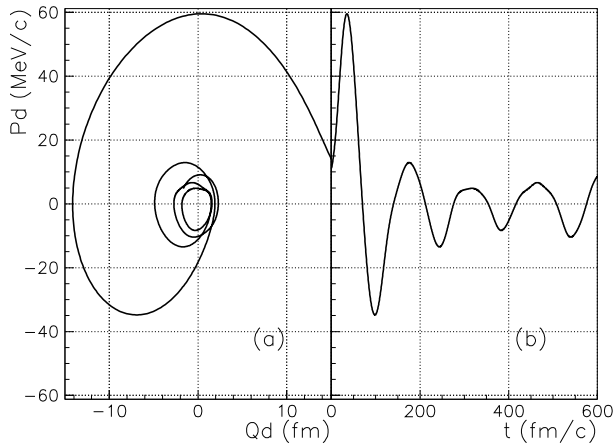}
\end{center}
\caption{Time evolution of the dipole vibration. Dipole moment $Q_d$ and its
conjugate $P_d$ are plotted in the phase space (a) and $P_d$ is plotted in
function of time (b).}
\label{fig:3}
\end{figure}

The period of the observed oscillations is around $150$ fm/c while for the $%
^{40}$Ca nucleus in its ground state it is almost half this value. This
large difference can be explained by the deformation of the fused system.
Indeed, in the TDHF simulations the compound nucleus only slowly relaxes its
initial prolate elongation along the axis of the collision. The averaged
value of the observed quadrupole deformation parameter is around $%
\varepsilon =0.23$. For symmetry reasons, the dipole oscillation occurs only
along the deformation axis of the nucleus formed by fusion in head on head
reactions. Therefore, a lower mean energy is expected for this longitudinal
collective motion according to the following relation $E_{GDR_{Z}}=E_{GDR}%
\left( 1-\varepsilon \right) ^{2}$ The energy of the GDR along the
elongation axis ${E_{GDR}}_Z$ fullfils this relation with $\varepsilon
\approx 0.26$ in a good agreement with the previous calculation of $%
\varepsilon$.

\subsection{Dynamical coupling of the GDR with the nucleus deformation}

To get a deeper insight in the dipole oscillation observed in fusion
reactions we have analyzed the time evolution of the period. From each point
on the collective trajectory in the collective phase space $
(Q_{_{d}},P_{_{d}})$ this quantity can be inferred from the time needed to
reach the opposite side of the observed spiral (see Fig. 3). The resulting
evolution is plotted in Fig. $4-a$. This period presents oscillations too.
These variations of the GDR period are almost in phase with the observed
oscillations of the monopole moment $Q_{0}$ and the quadrupole moment $Q_{2}$
presented in Fig. $4-b$. This points to a possible coupling between the
dipole mode and another mode of vibration. The evolutions of the monopole
and quadrupole moments are very similar. In particular, they present the
same oscillation's period around $166$ fm/c. Therefore, they originates from
the same phenomenon, the vibration of the density around a prolate shape.
This oscillation modifies the properties of the dipole mode in a time
dependent way.

\begin{figure}[tbh]
\begin{center}
\includegraphics[width=8cm]{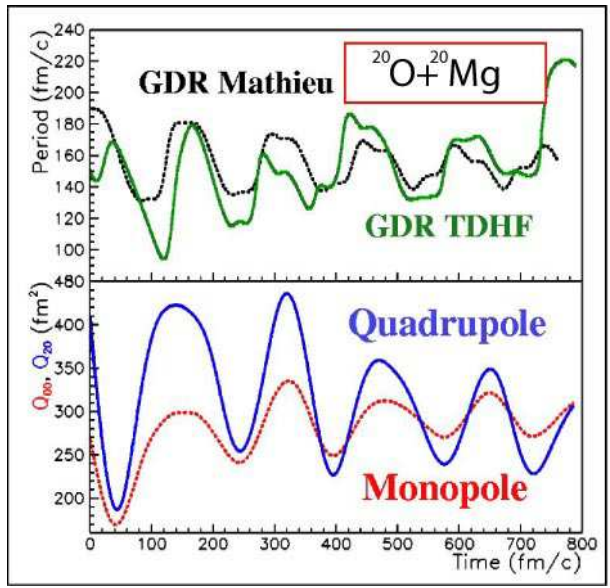}
\end{center}
\caption{\protect\smallskip (a) Time behaviour of the dipole period (solid
line) and its modelisation with the Mathieu's equation (dashed line). (b)
Time evolution of the monopole (dashed line) and quadrupole moments (solid
line).}
\label{fig:4}
\end{figure}

\smallskip

In order to investigate if this is the origin of the observed behavior we
can model the induced mode coupling. Let us consider a harmonic oscillator
of average frequency $\omega _{0}$ and let us replace the influence of the
density oscillation characterized by the frequency $\omega $ by assuming
that its spring constant varies in time at a frequency $\omega $. In such a
model, the equation of the motion becomes the Mathieu's equation 
\begin{equation}
\frac{\ddot{x}}{\omega _{0}^{2}}+[1+\delta \cos(\omega t)]x=0
\label{Eq:Mathieu}
\end{equation}
where $\omega _{0}$ is the pulsation without coupling and $\omega $ the
pulsation of the density's oscillation while $\delta$ corresponds to the
magnitude of the induced frequency fluctuations. We have computed the
numeric solution of Eq. \ref{Eq:Mathieu} with the typical external frequency
of the monopole and quadrupole oscillations. The bare frequency $ \omega _{0}
$ and the coupling strength $\delta$ have been tuned in order to get the
same oscillation frequency and typical amplitude. Indeed, because the
presence of the oscillating term the observed frequency is different from
the bare one. The solution (see Fig. 4-a) well reproduces our observations
with $\delta =-0.32$ and $\omega _{0}$ changed by a factor $1.2$ from the
observed value $E_{GDR_{Z}}$. From this analysis it appears that the
observed dipole motion corresponds to a giant vibration along the main axis
of a fluctuating prolate shape.

\subsection{Couplings between modes}

It is interesting to relate $\delta $ to a coupling matrix element between
collective states. Indeed, the Mathieu's equation can be seen as the
equation of motion coming from an Hamiltonian containing a coupling term
between the dipole mode and the collective deformation (GMR or GQR called
here after $\mu $) $V=\frac{k}{2}\delta \frac{1}{{\langle{Q}_{\mu
}\rangle_{max}}}Q_{\mu }Q_{_{d}}^{2}.$ This leads to a coupling matrix
element between the GDR and the state $\mu $ built on top of it which reads $
V=\delta \frac{\omega _{0}}{2}\frac{q_{\mu }}{{\langle{Q}_{\mu }\rangle_{max}%
}} B_{\mu }^{\dagger }B_{d}^{\dagger }B_{_{d}}$. In the studied case for
both the monopole and quadrupole, using the ground state matrix elements $%
q_{\mu },$ the amplitude of the oscillation $\frac{{\langle{Q}_{\mu
}\rangle_{max}}}{q_{\mu }}$ is about $5.$ Using $\omega _{0}\simeq 13$ MeV
we get a coupling between the dipole and the monopole or quadrupole $%
v=\delta \frac{\omega _{0}}{2}\frac{q_{\mu }}{{\langle{Q}_{\mu }\rangle_{max}%
}}=-0.5$ MeV. This value is qualitatively in agreement with the previous
observation of a strong coupling. From a quantitative point of view, it is
lower than the corresponding one (see table $1$). However, it is obtained
for a hot and deformed system with a large amplitude monopole and quadrupole
motion. Moreover the value derived here only correspond to the coupling with
a unique mode and so should rather be compared with table 2.

\section{GR in diluted nuclei}

In violent heavy ion collisions, nuclear matter is excited and compressed.
Then the formed nuclear system expands under the resulting thermal and
mechanical pressure. Matter may also be quenched in the coexistence region
of the nuclear liquid-gas phase diagram and the observed abundant fragment
formation may take place through a rapid amplification of spinodal
instabilities. New experimental results pleading in favor of such a spinodal
decomposition have recently been reported \cite{Beanliean,Borderie}. From
the theoretical point of view, the spinodal instabilities in finite systems
are unstable collective motions. They have been mainly studied within
semi-classical or hydrodynamical framework \cite
{Bertsch,Heiselberg,Norenberg,Colonna,Jacquot1}. However, since the relevant
temperatures are comparable to the shell spacing and the wave numbers of the
unstable modes are of the order of Fermi momentum, quantum effects are
expected to be important. Quantum approaches linking the spinodal
instabilities with the giant resonances can be found in \cite
{Ayik,Jacquot2,PRL-spino}.

\subsection{\protect\smallskip RPA in diluted systems}

To investigate instabilities encountered during the evolution of an
expanding system, one should study the dynamics of the small deviations $
\delta {\rho (t)}$ around the TDHF trajectory $\rho (t)$ \cite{Vautherin1}.
It is more convenient to carry out such an investigation in the ''co-moving
frame'' and if we consider the early evolution of instabilities in the
vicinity of an initial state the problem reduces to an RPA like equation 
\cite{PRL-spino}. Then, small density fluctuations are characterized by the
RPA modes $\rho _{\nu }$ and the associated frequencies $\omega _{\nu }$.
When the frequency of a mode drops to zero and then becomes imaginary, the
system enters an instability region.

In order to perform an extensive study of instabilities we may parametrize
the possible densities $\rho _{0}$ either by a static Hartree-Fock (HF)
calculations constrained by a set of collective operators \cite{Sagawa}, or
using a direct parameterization of the density matrix. In the following, we
follow the second approach by introducing a self-similar scaling of the HF
density as suggested by dynamical simulations.

First, we solve the HF equation for the ground state $\left[ h_{HF},\rho
_{HF}\right] =0$, leading to the single-particle wave functions $|\varphi
_{i}\rangle$ and the associated energies $\varepsilon _{i}$. Then we
introduce the density matrix at a finite temperature $T$ as $\rho
_{HF}\left[ T\right] =1/\left( 1+\exp \left( \left( h_{HF}\frac{\;}{\;}%
\varepsilon _{F}\left[ T\right] \right) /T\right) \right) $, where $%
\varepsilon _{F}\left[ T\right] $ is the corresponding Fermi level that is
tuned in order to get the correct particle number. Then, we perform a
scaling transformation, $R\left[ \alpha \right] ,$ which inflates the wave
functions in radial direction by a factor $\alpha $ according to 
$\langle r%
|R\left[ \alpha \right] |\varphi \rangle=\alpha ^{-1/3}\langle r/\alpha
|\varphi \rangle.$ We then define the density matrix for a hot and diluted
system by $\rho _{0}\left[ \alpha ,T\right] =R\left[ \alpha \right] \;\rho
_{HF}\left[ \alpha ^{2}T\right] \;R^{\dagger }\left[ \alpha \right] $. The
eigenstates of the constrained Hamiltonian are given by $|i\rangle=R\left[
\alpha \right] |\varphi _{i}\rangle$, and the corresponding energies and
occupation numbers are $\epsilon _{i}=\varepsilon _{i}/\alpha ^{2}$ and $%
n_{i}=1/\left( 1+\exp \left( \left( \varepsilon _{i} \frac{\;}{\;}%
\varepsilon _{F}\left[ \alpha ^{2}T\right] \right) /\alpha ^{2}T\right)
\right) $, respectively (see Ref. \cite{PRL-spino} for more details).

\begin{figure}[tbh]
\begin{center}
\includegraphics[width=.45\linewidth]{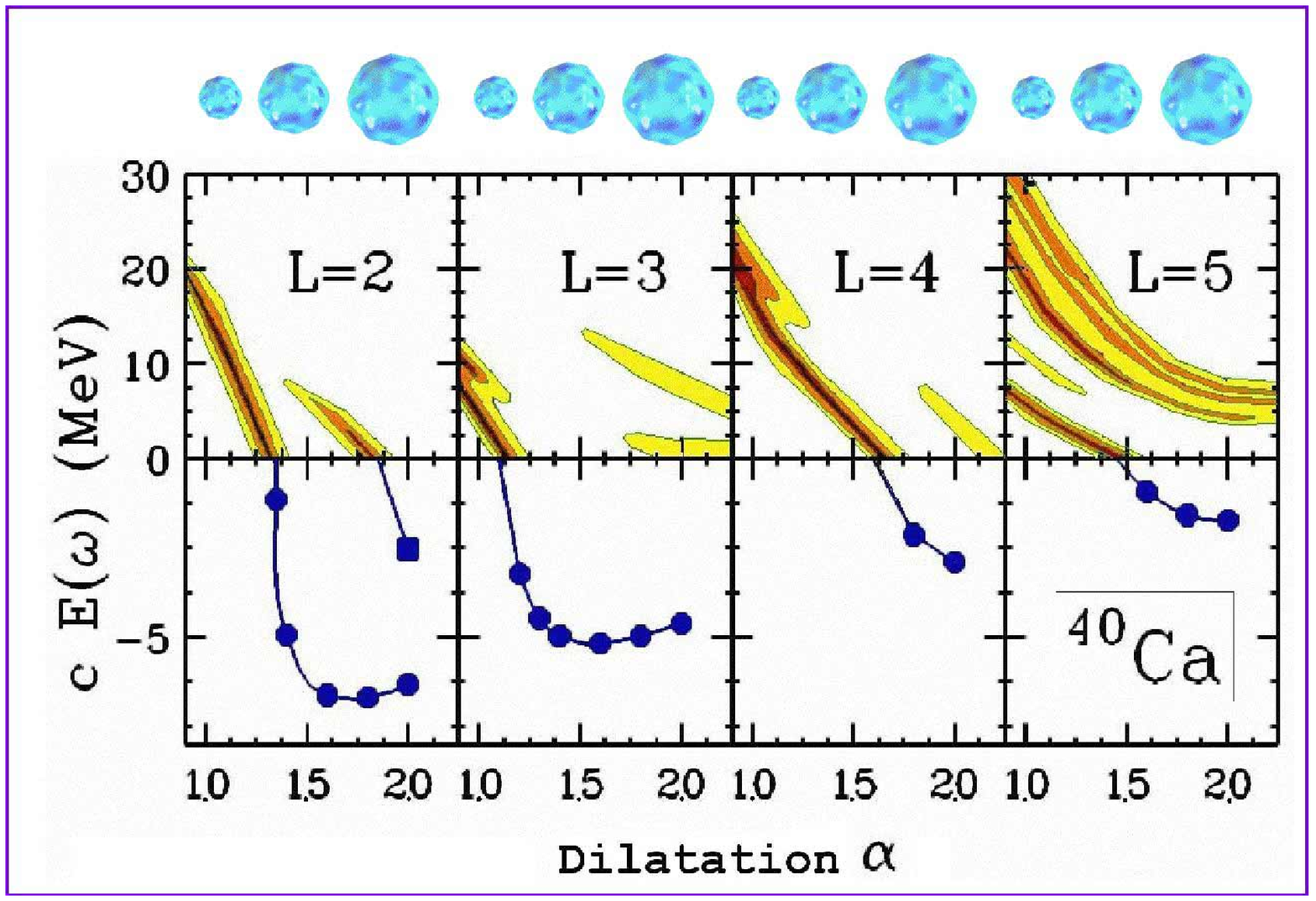} %
\includegraphics[width=.45\linewidth]{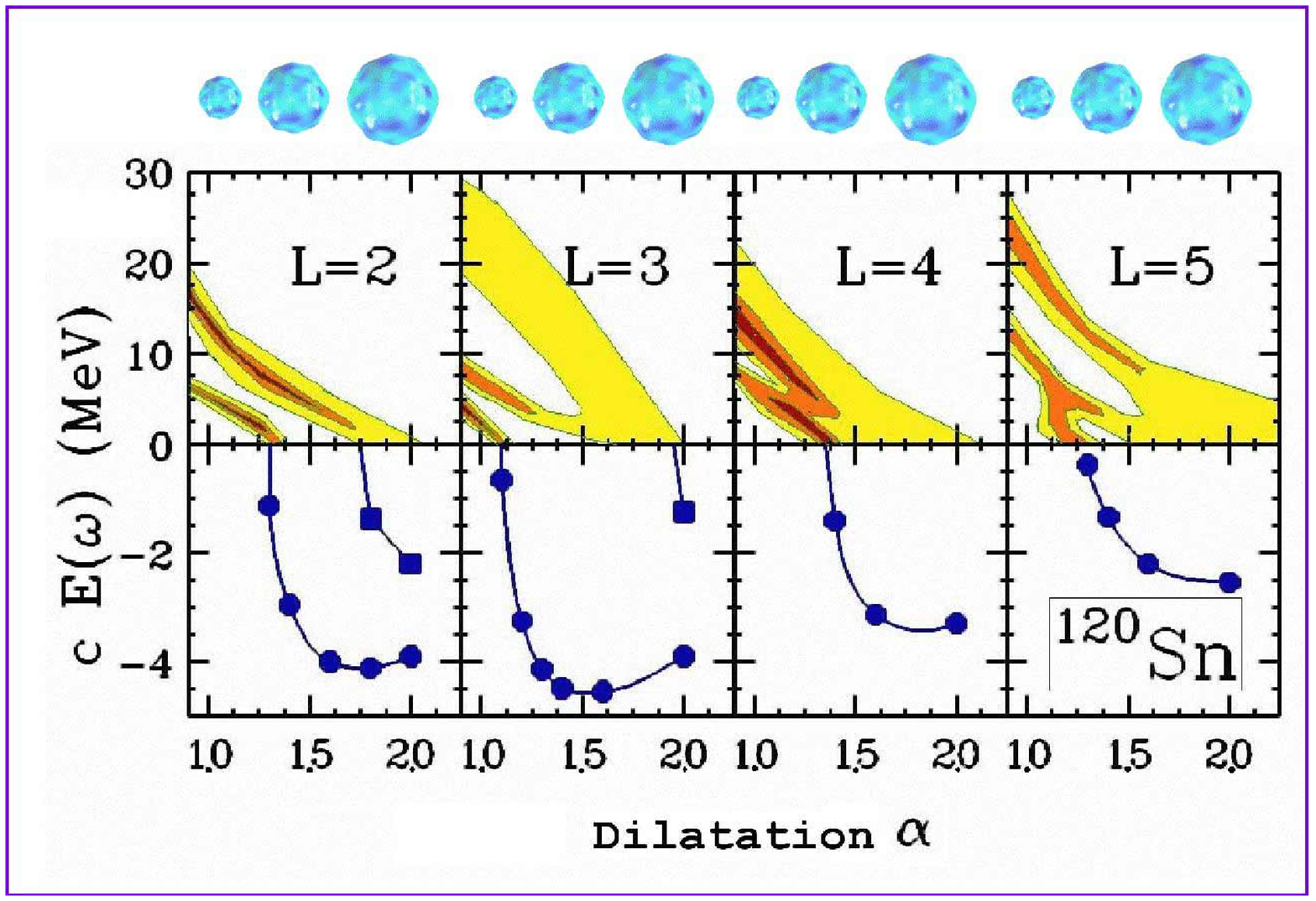}
\end{center}
\caption{Contour plots of the isoscalar strength functions associated with
the multipolarity $L=2-5$ as a function of the dilution parameter $\alpha $
and the collective energy of the mode $E_{\nu }=c\hbar \omega _{\nu }$ ($c=1$
for stable modes, $-i$ for unstable modes) for $^{40}$Ca (left) and $^{120}$%
Sn (right). }
\label{Fig1-RPA}
\end{figure}

We performed the HF\ calculations in the coordinate representation using the
Skyrme force SLy4 \cite{Chabanat}. The particle and hole states are obtained
by diagonalizing the HF Hamiltonian in a large harmonic oscillator
representation \cite{VanGiai}, which includes 12 major shells for Ca
isotopes and 15 for Sn. We apply the scaling and heating procedures
described above to the density matrix, and calculate the residual
interaction in a self consistent manner. We solve the RPA by a direct
diagonalization using a discrete two quasi-particle excitation
representation \cite{Vautherin2}.

\subsection{Dilution-dependent GR\ frequencies}

The top part of Fig. \ref{Fig1-RPA} shows calculations performed for $ ^{40}$%
Ca. Top panels shows contour plots of the isoscalar strength function
associated with the isoscalar operator $A_{LM}^{s}=
\sum_{i=1}^{A}r_{i}^{L}Y_{LM}$, with multipolarity $L=2-5$, as a function of
the dilution parameter $\alpha$. We observe that, in the stable domain, the
energy associated with the dominant isoscalar strength decreases as dilution
becomes larger, and at a critical dilution it drops to zero. At larger
dilution, the system becomes unstable, and for each multipolarity, one or
two unstable modes appear. This is illustrated in the bottom panel, where
the ''imaginary energy'' of the mode $E_{\nu }=-i\hbar \omega _{\nu }$ is
plotted as a function of the dilution. Looking at the RPA solution in the
coordinate space, it appears that the collective motions transform into
volume modes when they become unstable. We observe in fact a quite complex
structure of the unstable modes: Volume and surface instabilities are
generally coupled, as well as isoscalar and isovector excitations, since
protons and neutrons move in oposite way \cite{PRL-spino}.


\subsection{RPA-instabilities\ ''phase diagram''}

We, also, carry out calculations at finite temperature and determine the
dilutions at which different unstable modes begin to appear. This allows us
to specify the border of the instability region in the density-temperature
plane for different unstable modes. Fig. 4 shows phase diagrams for octupole
instabilities in $^{120}$Sn. 
\begin{figure}[tbh]
\begin{center}
\includegraphics[width=5cm]{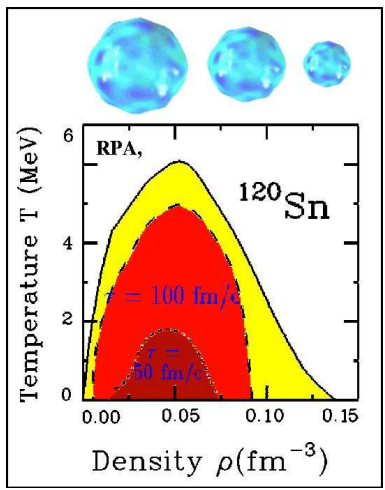}
\end{center}
\caption{ Border of the instability region (full fine) associated with $L=3$%
, for $^{120}$Sn. The dashed line connects the points having the instability
growth time $\tau $ = 100~fm/c. The dots are associated with $\tau $ =
50~fm/c. }
\end{figure}
Here, for simplicity, we define the density as $\rho =\rho _{0}/{\alpha }%
^{3} $. The full line indicates the border of the instability region. The
dashed line connects points that are associated with the instability growth
time $ \tau $ = 100~fm/c, and the dots correspond to situations with a
shorter growth time $\tau $ = 50~fm/c. We observe that in finite nuclei the
instability region is quite reduced as compared to that of nuclear matter.
The limiting temperature for instability to occur is around 6 MeV for Sn and
4.5 MeV for Ca (see \cite{PRL-spino}) while it is about $16$ MeV in
symmetric nuclear matter. As a result, heavier systems have larger
instability region than the lighter ones.

\subsection{Link with the coupling between modes}

The observed dilution dependence of the GR's energies can be interpreted in
terms of a coupling between the studied GR and the GMR. Indeed, it
corresponds to the Hamiltonian $H=\omega \left( \alpha \right)
B_{d}^{\dagger }B_{_{d}}$ where $\omega \left( \alpha \right) $ is the GR
frequency for a dilution $\alpha$. Using a Taylor expansion of $\omega
\left( \alpha \right) $ around $\alpha =1$ and introducing $\varepsilon
=\alpha -1,$ $\omega \left( \alpha \right) =\omega _{0}+\varepsilon \left.
\partial _{\alpha }\omega \right| _{0},$ and the Hamiltonian becomes $%
H=\omega _{0}B_{d}^{\dagger }B_{_{d}}+\varepsilon \left. \partial _{\alpha
}\omega \right| _{0}B_{d}^{\dagger }B_{_{d}}.$ The dilution factor $\alpha $
can be related to the collective observables using $\langle r%
^{2}\rangle=\alpha ^{2}\langle r^{2}\rangle_{0}\simeq \langle r%
^{2}\rangle_{0}+2\varepsilon \langle r^{2}\rangle_{0}$ and $Q_{\mu }= \frac{1%
}{\sqrt{4\pi }}\left( r^{2}-\langle r^{2}\rangle_{0}\right) =q_{\mu }\left(
B_{\mu }^{\dagger }+B_{\mu }\right) $ so that the Hamiltonian reads $
H=\omega _{0}B_{d}^{\dagger }B_{_{d}}+\left[\frac{\sqrt{4\pi }q_{\mu }}{2%
\langle r^{2}\rangle_{0}}\left. \partial _{\alpha }\omega \right| _{0}B_{\mu
}^{\dagger }B_{d}^{\dagger }B_{_{d}}+h.c.\right]$. This leads to a coupling
matrix element between the GR and a GMR ($\mu $) built on top of it which
reads $V=\frac{\sqrt{4\pi }q_{\mu }}{2\langle r^{2}\rangle_{0}}\left.
\partial _{\alpha }\omega \right| _{0}B_{\mu }^{\dagger }B_{d}^{\dagger
}B_{_{d}}$. Looking at the GQR decrease with the dilution, we get $\partial
_{\alpha }\omega \simeq -40$ MeV for $^{40}$Ca. Using the transition matrix
element for the GMR extracted in table 1, $q_{\mu }\simeq 12$ fm$^2$, and $%
\langle r^{2}\rangle_{0}/\sqrt{4\pi } \simeq 125$ fm$^2$ we get $v_\mu=\frac{%
q_{\mu }}{2\langle r^{2}\rangle_{0}/\sqrt{4\pi }}\left. \partial _{\alpha
}\omega \right| _{0}$ $\simeq -2$ MeV. This value is qualitatively in
agreement with the previous observation of a strong coupling. In particular
we understand the sign of the interaction since the energy of the modes is
reduced in a diluted system ($v_\mu<0$). From a quantitative point of view,
it is a factor 2 lower than the one reproduced in table $1$. However one
should remember that we compute here an average matrix element while in
table 1 it is more an integrated one. In fact this result should be compared
to the one in table 2 and indeed they are in excellent agreement.

\smallskip

\section{\protect\smallskip Conclusion}

In conclusion, we have presented a study of 3 different situations: the
excitation of small amplitude motions and the coupling between phonon
states, the excitation of pre-equilibrium GDR in fusion reaction and the
modification of the GDR properties due to the dynamics of the monopole and
quadrupole deformation and finally the early development of spinodal
instabilities. The 3 phenomena show that collective modes are strongly
coupled to the monopole (and quadrupole) degrees of freedom. We have
presented original methods to extract the coupling matrix element between a
collective mode and a GMR (or GQR) built on top of it. The 3 studied
phenomena are in qualitative agreement pointing to a negative coupling of
the order of the MeV. The quantitative difference we observe which might
come the differences between the 3 studied cases is now under study.

\end{document}